\documentstyle[preprint,aps,epsf]{revtex}
\tighten
\draft
\widetext
\input epsf
\preprint{UPR-978-T}
\bigskip
\bigskip
\begin{document}
\title{\large \bf On
the Signature of Short Distance Scale in the \\
\vspace{1mm}
Cosmic Microwave Background}
\medskip
\author{
Gary Shiu$^{1}$ and Ira Wasserman$^{2}$  \vspace{0.2cm}}
\address{$^1$ Department of Physics and
Astronomy, University of Pennsylvania, Philadelphia, PA 19104 \\
$^2$ Center for Radiophysics and Space Research, Cornell University, Ithaca, NY 14853}
\date{March 12, 2002}
\bigskip
\medskip
\maketitle

\def\kl#1{\left(#1\right)}
\def\th#1#2{\vartheta\bigl[{\textstyle{  #1 \atop #2}} \bigr] }

\begin{abstract}
{} We discuss the signature of the scale of short distance physics
in the Cosmic Microwave Background. In addition to effects which
depend on the ratio of Hubble scale $H$ during inflation to the
energy scale $M$ of the short distance physics, there can be effects
which depend on $\dot{\phi}^2/M^4$ where $\phi$ is the {\it
classical background} of the inflaton field. Therefore, the
imprints of short distance physics on the spectrum of Cosmic
Microwave Background anisotropies generically involve a {\it
double expansion}. We present some examples of a single scalar
field with higher order kinetic terms coupled to Einstein gravity,
and illustrate that the effects of short distance physics on the
Cosmic Microwave Background can be substantial even for $H << M$,
and generically involve corrections that are not simply powers of
$H/M$. The size of such effects can depend on the short distance
scale non-analytically even though the action is local.
\end{abstract}

\def\kl#1{\left(#1\right)}
\def\th#1#2{\vartheta\bigl[{\textstyle{  #1 \atop #2}} \bigr] }
\newcommand{\drawsquare}[2]{\hbox{%
\rule{#2pt}{#1pt}\hskip-#2pt
\rule{#1pt}{#2pt}\hskip-#1pt
\rule[#1pt]{#1pt}{#2pt}}\rule[#1pt]{#2pt}{#2pt}\hskip-#2pt
\rule{#2pt}{#1pt}}

\newcommand{\fund}{\raisebox{-.5pt}{\drawsquare{6.5}{0.4}}}
\newcommand{\Ysymm}{\raisebox{-.5pt}{\drawsquare{6.5}{0.4}}\hskip-0.4pt%
        \raisebox{-.5pt}{\drawsquare{6.5}{0.4}}}
\newcommand{\Yasymm}{\raisebox{-3.5pt}{\drawsquare{6.5}{0.4}}\hskip-6.9pt%
        \raisebox{3pt}{\drawsquare{6.5}{0.4}}}
\newcommand{\antifund}{\overline{\fund}}
\newcommand{\bYasymm}{\overline{\Yasymm}}
\newcommand{\bYsymm}{\overline{\Ysymm}}




\section{Introduction}

One of the highlights of inflationary cosmology \cite{Guth,new} is the idea
that astrophysical
observables may be the results of physics at the microscopic scale.
Since the pioneering work of \cite{fluctuations},
it has been known that quantum field
fluctuations in the early universe are stretched by the enormous
expansion of inflation to scales of astrophysical size, providing
a first principle mechanism for density perturbations.
In an inflationary universe,
the temperature anisotropies of the Cosmic Microwave Background (CMB)
and even the formation of galaxies
are the results of quantum fluctuations writ large.

Most inflationary models have far more expansion than the 60 e-foldings
required to solve the cosmological problems of standard cosmology.
Consequently, astrophysical scales in the present universe
may correspond to Planckian or sub-Planckian scales at the onset of
inflation. Therefore, there is a hope that
inflation may provide a kind of Planck scale microscope,
magnifying short distance physics to observably large scale,
allowing us to
probe trans-Planckian or
stringy physics from precision cosmological
measurements \cite{brandenberger,niemeyer,CGS,EGKS1,EGKS2,ShiuTye,Hui}.

More recently, this exciting prospect was revisited in Ref.\cite{Shenker},
where it was claimed that a general argument assuming only low energy
locality implies that the effects of
short distance physics on the CMB are of the order of
$(H/M)^{2n}$ where $n \geq 1$ is an integer,
$H$ is the Hubble scale during inflation, and $M$ is
the mass scale of short distance physics.
This result seems to imply that stringy effects are far too small to
be observed.
However, it was pointed out in \cite{NewBrandenberger}
that the argument of \cite{Shenker} assumes
the universe is in the local
vacuum state at horizon crossing, and
hence the result of \cite{Shenker} is not in
conflict with the seemingly different estimates
of \cite{EGKS1,EGKS2,BrandenbergerHo}.

The purpose of this paper is to point out yet another way that the
scale of short distance physics can affect the spectrum of CMB
anisotropies.
The effects of short distance physics ({\it e.g.}, stringy
$\alpha^{\prime}$ corrections) on the low energy effective action
of a scalar field
 generically
include both higher derivative terms as well as higher order terms
in the first derivative. Therefore, the effects
of short distance physics on the CMB anisotropies is a double
expansion in both $(H/M)^2$ and $\dot{\phi}^2/M^4$ where $\phi$ is
the {\it classical background} of the scalar field\footnote{The
quantum fluctuation $\delta \phi$ obeys a linear equation, and so
the action contains at most terms quadratic in $\delta \phi$.
Therefore, the remaining factors in these higher order terms are
evaluated on the classical background.}. Contrary to the
assertions of \cite{Shenker}, the imprint of the short distance
scale on the CMB generically involve corrections that are not
simply powers of $(H/M)^2$. We find that the effects of short
distance physics can be substantial even though $H<<M$, and can in
fact depend on the mass scale of short distance physics {\it
non-analytically} even though the action is {\it local}.

The effects of higher derivative terms on the CMB anisotropies
have been studied in
\cite{brandenberger,niemeyer,CGS,EGKS1,EGKS2,Shenker}. Here we
focus on the effects of the higher order terms in the first
derivative. Let us consider the most general local action for a
scalar field coupled to Einstein gravity, which involves at most
first derivatives of the scalar field:
\begin{equation}
S = - \frac{1}{16 \pi G} \int d^4 x \sqrt{g} R +  \int d^4 x \sqrt{g}~ p(X, \phi)
\label{lagrangian}
\end{equation}
where $g$ is the determinant of the metric, $R$ is the Ricci scalar and
\begin{equation}
X= \frac{1}{2} g^{\mu \nu} \partial_{\mu} \phi \partial_{\nu} \phi~.
\end{equation}
The cosmological perturbations of this class of models
have been studied in \cite{Garriga}.
In this paper, we apply their results to some special cases to illustrate 
that the short distance effects on
the Cosmic Microwave Background may not be captured
simply by an expansion
in $(H/M)^2$, and
can depend on the short distance energy
scale $M$ in a non-analytic way even though
the action which defines the model involves only local interactions.

The Lagrangian of the scalar field is denoted by $p(X,\phi)$ since
it plays the role of
pressure \cite{ADM},
whereas the energy density is given by
\begin{equation}
\varepsilon = 2 X p_{,X} - p
\end{equation}
where $p_{,X}$ denotes the derivative of the Lagrangian with respect
to $X$.

We consider the function $p(X,\phi)$ of the following form:
\begin{equation}
p(X,\phi) = F(X) -V(\phi)
\label{pdef}
\end{equation}
and hence,
\begin{equation}
\varepsilon = 2 X F_{,X} - F + V~.
\end{equation}
The function $F(X)$ includes the usual kinetic term $X$ as well
as higher order contributions.
Therefore,
\begin{equation}
F(X) = X + \alpha X f \left({X\over M^4}\right)
\label{fdef}
\end{equation}
where $\alpha$ is a dimensionless number,
$f$ is a function of $X/M^4$, and
$M$ is the mass scale associated with the short distance physics.
If $\alpha=0$, we have the usual kinetic term.
If $\alpha \not= 0$,
the kinetic term is not minimal and the higher order kinetic terms
depend on $M$.

%
The rationale behind Eq. (\ref{pdef}), augmented by Eq. (\ref{fdef}),
is that it is a plausible prescription for encoding the physics
of a small distance scale, $M$, in a way that depends rather simply on
fields and first derivatives of the fields, and can preserve causality if
appropriate conditions on $F(X)$ are satisfied. Dimensional analysis
alone would allow other choices, such as
\begin{equation}
p(X,\phi)=XG_K\left({X\over\phi^2M^2}\right)-V(\phi)
G_V\left({X\over\phi^2M^2}\right)~;
\label{otherchoice}
\end{equation}
in some inflationary scenarios, a theory based on Eq. (\ref{otherchoice})
could result in corrections to conventional inflation that can
be expressed as a power series in $H^2/M^2$ for small $H/M$, as
envisioned in \cite{Shenker}. For example, in the chaotic inflationary scenario
developed in \S~\ref{chaotic}, the dimensionless combination
$X/\phi^2M^2\approx H^2/8M^2L^2$ during slow roll, where $L$ is the
number of e-folds remaining to the end of inflation.
However, we shall see that the simpler prescription based on 
Eqs. (\ref{pdef}) and (\ref{fdef}) can lead to more complicated
behavior.
%

\section{General Results for Slow-roll inflation}

\subsection{Background Equations}

We consider the background to be an expanding Friedmann universe:
\begin{equation}
ds^2 = dt^2 - a^2(t) \delta_{ij} dx^i dx^j~.
\end{equation}
The equations of motion for the background variables $a(t)$ and $\phi (t)$
are:
\begin{eqnarray}
H^2 &=& \frac{8 \pi G}{3} \varepsilon ~, \nonumber \\
\dot{\varepsilon} &=& -3 H \left( \varepsilon + p \right)~.
\end{eqnarray}
In the slow-roll approximation,
\begin{eqnarray}
|2X F_{,X} - F| ~&<<& ~V \nonumber \\
| \left(2 X F_{,X} - F \right)_{,X} \ddot{\phi}| ~&<<&~ \frac{\partial V}{\partial \phi}~,
\label{slowrollconds}
\end{eqnarray}
the background equations become
\begin{eqnarray}
H^2 & \approx & \frac{8 \pi G}{3} V\nonumber \\
3 H \dot{\phi} F_{,X} &\approx& - \frac{\partial V}{\partial \phi}~.
\end{eqnarray}

\subsection{Fluctuation Spectra}

The power spectrum for the scalar fluctuations is given by \cite{Garriga}
\begin{equation}
P_k^{S} = \left. \frac{16}{9} \frac{G^2 \varepsilon^2}{c_s \left( \varepsilon + p\right)} \right|_{c_s k = a H}
\end{equation}
where the quantities on the right-handed side are evaluated at
sound horizon crossing, {\it i.e.}, $aH=c_s k$.
Here, $c_s$ is the ``speed of sound'' for the scalar perturbations:
\begin{equation}
c_s^2 = \frac{\varepsilon + p}{2 X \varepsilon_{,X}}
= \frac{F_{,X}}{F_{,X} + 2 X F_{,XX}} ~.
\label{soundef}
\end{equation}
Causality requires that $c_s \leq 1$ which implies that
$2X F_{,XX}/F_,X \geq 0$.

In the slow-roll approximation
\begin{equation}
P_k^S \approx \left. \frac{16}{9} \frac{G^2 V^2}{c_s \dot{\phi}^2 F_{,X}}
 \right|_{c_s k = a H}
= \left. \frac{16}{9} \left( \frac{3}{8 \pi} \right)^2
\frac{H^4}{(c_s F_{,X}) \dot{\phi}^2}  \right|_{c_s k = a H}~,
\label{scalarflucts}
\end{equation}
where the quantities are evaluated at ``sound horizon'' crossing,
but,
on the other hand, the spectrum of tensor fluctuations is given by
the usual expression.
\begin{equation}
P_k^T = \left. \frac{128}{3} G^2 \varepsilon
\right|_{k = a H}
\approx \left. \frac{128}{3} G^2 V \right|_{k = a H}
= \left. \frac{16}{\pi} G H^2 \right|_{k = a H}~,
\end{equation}
where the quantities are evaluated at the usual particle horizon crossing,
$k=aH$.
This is not exactly the same time as the sound horizon crossing
for the scalar perturbations but as we shall see, in the specific model that
we consider, $c_s\sim 1$, so the two horizon crossing times differ
by relatively small amounts.

Note that the short distance effects are contained in the function
$F(X)$ (and its derivatives) evaluated at horizon crossing.
Therefore the short distance effects on the perturbation spectra
should depend on $\dot{\phi}^2/M^4$.

\section{Examples}

\subsubsection{Chaotic Inflation Model}
\label{chaotic}

Let us consider the potential in chaotic inflation \cite{chaotic}:
\begin{equation}
V(\phi) = \frac{1}{2} m^2 \phi^2~.
\end{equation}
The background equations are especially simple:
\begin{eqnarray}
H & \approx& \sqrt{\frac{4 \pi G}{3}} m \phi \nonumber \\
F_{,X} \dot{\phi} &\approx& - \frac{m}{\sqrt{12 \pi G}}~.
\label{evolvemodel}
\end{eqnarray}
Hence, $\dot{\phi}$ and $F_{,X}$ are independent of time.
We can express $\phi$ and $H$ as functions of the number of e-foldings
remaining to the end of inflation, {\it i.e.}, $L=\ln (a_{end}/a)$:
\begin{eqnarray}
\phi^2 &=& \frac{L}{2 \pi G F_{,X}} \nonumber \\
H^2 &=& \frac{2 m^2 L}{3 F_{,X}}~.
\label{solvemodel}
\end{eqnarray}
Therefore, the scalar and tensor perturbations are
\begin{eqnarray}\label{power}
P_k^S &=& \frac{4 G m^2 L_S^2}{3 \pi}
\left[  \frac{ \left( F_{,X} + 2 X F_{,XX} \right)^{1/2}}{F_{,X}^{3/2}} \right]
\nonumber \\
P_k^T &=& \frac{32}{3 \pi} \frac{G m^2 L_T}{F_{,X}}~.
\end{eqnarray}
The usual results are recovered if we take $F(X)=X$. Since $L_s-L_T\simeq
{1\over 2}\ln c_s^{-1}$, and $c_s^2\geq [1+2(n-1)]^{-1}$ if $F(X)$ is a
polynomial of index $n$, the fractional difference $(L_S-L_T)/L_T\ll 1$
for large $L_T$, so we neglect $L_S-L_T$ below.
From the equation of motion:
\begin{equation}
F_{,X} \sqrt{2X} = - \frac{m}{\sqrt{12 \pi G}}
\label{Xeqn}
\end{equation}
We see that the solution of $X$ is, in general, not analytic
in $M$.

For example, consider
\begin{equation}
F(X) = X + \frac{\alpha}{M^4} X^2~,
\end{equation}
where $M$ is the mass scale characteristic of short distance physics, and
we introduce a parameter $\alpha$ which we may absorb into the definition
of $M$, but we carry along anyway to keep track of nonlinear terms in
$F(X)$. (If such terms are present, we take $\alpha=1$ below;
$\alpha=0$ means these terms are absent altogether.) With this $F(X)$, we find
\begin{eqnarray}
P_k^S &=& \frac{4 G m^2 L^2}{3 \pi}
\left[  \frac{ \left( 1 + 6 {\alpha}X/{M^4} \right)^{1/2}}{(1+2\alpha X/M^4)^{3/2}} \right]
\nonumber \\
P_k^T &=& \frac{32}{3 \pi} \frac{G m^2 L}{(1+2 {\alpha}X/{M^4})}
\end{eqnarray}
where $X$ is given by the solution of the cubic equation
\begin{equation}
\left( 1+ 2 \frac{\alpha}{M^4} X \right) \sqrt{2 X} =
\frac{m}{\sqrt{12 \pi G}}~.
\end{equation}
If $\alpha X/M^4\ll 1$, then $X$ is independent of $M$ to lowest
order, and so, to the same order, the power spectra are also independent
of $M$.
There are higher order corrections amounting to an expansion in $X/M^4$; in
this sense, the power spectra depend on $M$ analytically at small $X/M^4$.
However, if $\alpha X/M^4\gg 1$, then $X\approx
m^{2/3}M^{8/3}/(12\pi G)^{1/3}$, and therefore $P_k^{S,T} \sim
Gm^2M^4/X\sim G^{4/3}m^{4/3}M^{4/3} =(M/M_{Pl})^{8/3}(m/M)^{4/3}$,
where $M_{Pl}=G^{-1/2}$ is the Planck mass. Thus, the fluctuation
amplitude depends on a fractional power of the mass scale
associated with the short distance physics.
Below we shall see that if we want $m\sim M$ (or perhaps $m\lesssim M$
but not $m\ll M$), then we will be forced to the large $X/M^4$ limit.
In this case, it will be possible to have small $H/M$, but we shall also
see that $m\sim M\ll M_{Pl}$.

 The spectral index for the tensor fluctutations is
given by \cite{Garriga}:
\begin{equation}
n_T = - \frac{3 \left( \varepsilon + p \right)}{\varepsilon}
\approx - \frac{6 X F_{,X}}{V}
= - \frac{1}{L}
\end{equation}
Apart from the small change in $L$, the spectral index is not very different
from the usual case. Furthermore, since $c_s$ is of order $1$, the
``consistency condition'' is only altered mildly:
\begin{equation}
\frac{P^T_k}{P^S_k} = - 8 c_s n_T~;
\label{consistency}
\end{equation}
for the quadratic $F(X)$ adopted above, $c_s\geq 1/\sqrt{3}$.

So far, we have merely assumed that it is possible for the short
distance physics to alter the kinetic energy of the scalar field
enough to change both the expansion rate during inflation, and the
fluctuations that result. We have seen that the main effect would
be that the dependence of the fluctuation amplitude on mass scales
of the theory is changed, and the relative amplitudes of the
scalar and tensor fluctuations are altered by factors which, as
long as $c_s$ is not too small, are of order unity in general. In
the specific model we have considered, the relative amplitude of
the scalar and tensor fluctuations changes by a factor of at most
$\sqrt{3}\approx 1.73$ relative to its value for $\alpha=0$, a
subtle effect that may be discernible observationally,
particularly once polarization measurements become available for
CMB fluctuations. This change is model dependent, though, and
alternative $F(X)$ could yield different results, since $c_s$
could be smaller for large $X$. For example, $c_s\to M^4/2X$ for
large $X/M^4$ if $F(X)=M^4(e^{X/M^4}-1)$.
In this case, the
"consistency condition" is violated significantly and may give
rise to an observable effect.

An important question we have not asked is whether the large $X$
limit can be attained in any realistic inflation model. There are
three different issues to be addressed here:
\begin{itemize}
\item Is the large $X/M^4$ limit consistent within the context
of slow roll inflation?
\item Under what circumstances do we expect large $X/M^4$ to
arise? Are there any constraints on the mass scales of
the theory implicit in
the large $X/M^4$ limit?
\item Is the large $X/M^4$ limit consistent with the idea of
``integrating out'' massive modes in that the length scales
probed by inflation are larger than $M^{-1}$ even though
$X/M^4$ is large?
\end{itemize}
To examine these issues, let us consider the chaotic inflation
model more quantitatively.

First, let us check that the slow roll condition is not violated
in the limit of large $X/M^4$. With the help of Eqs. (\ref{evolvemodel})
and (\ref{solvemodel}) we see that $\vert\dot\phi\vert/H\phi\approx
(2L)^{-1}$, so that the scalar field evolves slowly compared with the
expansion rate of the Universe for all $L\gtrsim 1$, just as in
conventional chaotic inflation. (Notice that this particular condition
does not depend on $X/M^4$ at all.) From Eq. (\ref{slowrollconds}) we also
see that the energy density of the Universe is dominated by $V(\phi)$
until
\begin{equation}
2 X F_{,X} - F(X) \simeq V = \frac{m^2 L}{4 \pi G F_{,X}}~;
\end{equation}
where $V$ has been evaluated using the slow roll solution. This implies an
end to the slow roll phase when
\begin{equation}
L_{end} \sim \frac {1}{3} \left( 1 - \frac{F(X)}{2X F_{,X}} \right)
\end{equation}
which implies $L_{end}\sim 1$ whether or not $X/M^4$ is large. Finally,
in this particular cosmological model, $\ddot\phi=0$ during the slow
roll phase, so we do not have to consider additional constraints from
the smallness of $\ddot\phi$ as long as the other conditions underlying
the slow roll approximation are satisfied. Consequently, we see that
the conditions for slow roll to be valid are hardly affected by
the magnitude of $X/M^4$.

To assess the conditions under which large $X/M^4$ are likely to arise,
we need to find the mass scales implied in this regime. We shall see
that large $X/M^4$ is {\it required} if $m\sim M$, but not otherwise.

To make quantitative estimates, we use the fact that the amplitude
of scalar perturbations is fixed by observations of the CMB to be
$P_k^S\sim 10^{-10}$ on length scales for which $L\approx 60$.
If we define $u^2=2X/M^4$, then we find
\begin{equation}
{m\over M_{Pl}}={(3\pi P_k^S)^{1/2}\over 2L}\left[{(1+u^2)^{3/4}\over(1+3u^2)^{1/4}}
\right]\approx 2.6\times 10^{-7}\left(10^{10}P_k^S\right)^{1/2}
\left({60\over L}\right)\left[{(1+u^2)^{3/4}\over(1+3u^2)^{1/4}}
\right];
\label{meqn}
\end{equation}
for large values of $u$ this becomes
\begin{equation}
{m\over M_{Pl}}\approx 1.9\times 10^{-7}\left(10^{10}P_k^S\right)^{1/2}
\left({60\over L}\right)u.
\label{mbig}
\end{equation}
Thus, we see that $m\propto u$ at large $u$.

The value of the other mass scale associated with new physics can be
found by combining Eq. (\ref{meqn}) with Eq. (\ref{Xeqn}); the result
is
\begin{equation}
{M\over M_{Pl}}={(P_k^S)^{1/4}\over 2\sqrt{L} u^{1/2}[(1+u^2)(1+3u^2)]^{1/8}}
\approx {2.0\times 10^{-4}\left(10^{10}P_k^S\right)^{1/4}
\left({60/L}\right)^{1/2}\over u^{1/2}[(1+u^2)(1+3u^2)]^{1/8}};
\label{Meqn}
\end{equation}
for large values of $u$,
\begin{equation}
{M\over M_{Pl}}\approx 1.8\times 10^{-4}\left(10^{10}P_k^S\right)^{1/4}
\left({60\over L}\right)^{1/2}u^{-1}.
\label{Mbig}
\end{equation}
Thus, we see that $M\propto u^{-1}$ at large $u$.

The ratio of the two mass scales is
\begin{equation}
{m\over M}={\sqrt{3\pi}(P_k^S)^{1/4}\over\sqrt{L}}
\left[{(1+u^2)^{7/8}u^{1/2}\over (1+3u^2)^{1/8}}\right]
\approx 1.3\times 10^{-3}\left(10^{10}P_k^S\right)^{1/4}
\left({60\over L}\right)^{1/2}
\left[{(1+u^2)^{7/8}u^{1/2}\over (1+3u^2)^{1/8}}\right].
\end{equation}
Thus we see that if the mass scales are comparable, we must have large
values of $u$. In the large $u$ limit, we would have
\begin{equation}
{m\over M}\approx 1.1\times 10^{-3}\left(10^{10}P_k^S\right)^{1/4}
\left({60\over L}\right)^{1/2}u^2,
\end{equation}
and we can solve this to find
\begin{equation}
u\approx 30\left(10^{10}P_k^S\right)^{-1/8}\left({60\over L}\right)^{-1/4}
\left({m\over M}\right)^{1/2}.
\end{equation}
We can rewrite the two masses in terms of $m/M$, assuming large values of $u$:
\begin{equation}
{m\over M_{Pl}}\approx 5.9\times 10^{-6}(10^{10}P_k^S)^{3/8}\left({60\over L}\right)^{3/4}
\left({m\over M}\right)^{1/2},
\end{equation}
and $M=m/(m/M)$. When $m$ and $M$ are comparable, the large $u=\sqrt{2X/M^4}$ limit
{\it must} apply, and the relevant mass scale is well below $M_{Pl}$ i.e about
$7\times 10^{13}$ GeV.

If we insist that $M\sim M_{Pl}$, no inconsistencies arise, but we must then
require that $u\ll 1$, and therefore that $m\ll M$. In this case, we find
that $m/M_{Pl}\approx 2.56\times 10^{-7}\left(10^{10}P_k^S\right)^{1/2}
(60/L)$, or a mass scale $m\approx 3\times 10^{12}$ GeV, about a factor of
20 smaller than is found if $m\sim M$.
A theory with $M\sim M_{Pl}\gg m$ is {\it consistent}, and yields the conventional
chaotic inflation picture, modulo correction terms that may be expressed as
an expansion in $X/M^4\simeq  m^2M_{Pl}^2/24\pi M^4\simeq H^2M_{Pl}^2/16\pi M^4L$,
which is $\sim H^2/M^2$ for $M\sim M_{Pl}$, as was suggested in \cite{Shenker}.
However, not only is this behavior not required, but it is also strongly violated
if $m\sim M$. Moreover, even if $X/M^4\ll 1$ and $m\ll M$, the expansion parameter
$X/M^4$ generally involves two dimensionless ratios, $H/M$ and $M/M_{Pl}$, not
just $H/M$, and also depends on the number of e-folds remaining in inflation, $L$.

Finally, let us consider the third consistency issue, namely, whether the
length and time scales involved during inflation are all long compared with
$M^{-1}$, the fundamental scale of the theory. A truly complete treatment of
this question is beyond the intended scope of this paper, but a necessary
condition must be that $H/M$ be small. This condition must be satisfied
in order to justify using an effective theory, such as Eq. (\ref{lagrangian}),
for computing the small scale fluctuations that arise from quantum effects
during slow roll. If the inequality $H/M<1$ is satisfied, the use
of an effective theory for computing the evolution of the smooth background
is justified automatically, since, in slow roll, $\vert\dot\phi/H\phi\vert\ll 1$,
so that $\vert\dot\phi/M\phi\vert\ll H/M$.
In the limit of large $X/M^4$, we find that
\begin{equation}
{H\over M}\approx 0.21(10^{10}P_k^S)^{1/8}
\left({L\over 60}\right)^{1/4}\left({m\over M}\right)^{1/2}.
\end{equation}
If $m\lesssim M$, then the inequality $H<M$ ought to be satisfied, but not
very strongly. (We note that, by contrast, the limit $H/M_{Pl}\ll 1$ is satisfied
easily irrespective of the value of $X/M^4$.)
One may ask if we can trust an effective theory such as
Eq.(\ref{lagrangian}) when $X/M^4$ is large, since the higher
order corrections to $F(X)$ may be important. However, our
purpose here is to establish that there {\it can} be substantial
effects even when the length and time scales involved during inflation
are large compared with $M^{-1}$.
Thus, we expect that a treatment of early Universe cosmology
based on an effective theory such as Eq. (\ref{lagrangian}) ought to be
justified, but a more careful treatment than we have attempted here
is needed to establish this point rigorously.


\subsubsection{Other Potentials}

Although we do not consider other inflation potentials in detail here,
we should examine whether the behavior we have found is specific to
inflation in the chaotic inflation potential $V(\phi)={1\over 2}m^2\phi^2$.
To investigate this, let us ask
whether conventional inflation would lead to large values of $X/M^4$
for other potentials, taking account of constraints on the scalar fluctuation
amplitude.
Let us
consider instead potentials that are very flat functions of $\phi$, so
that $V=V_0[1+f(\phi/\phi_0)]\simeq V_0$ throughout slow roll
inflation. Then, from the validity of the slow roll approximation,
we know that $\dot\phi^2/M^4\ll V_0/M^4$; if $V_0\lesssim M^4$,
then the small $X/M^4$ limit is guaranteed. In this case, we do expect
the corrections to conventional inflation to be small, but they still
need not be expressible simply in terms of $H/M$. For example, if
$V(\phi)=V_0(1-e^{-\phi/\phi_0})$, we find that
\begin{equation}
{X\over M^4}={H^2M_{Pl}^4\over 128\pi^2\phi_0^2M^4
(1+L/8\pi G\phi_0^2)^2}~,
\end{equation}
and if $V=V_0[1-(\sigma+1)^{-1}(\phi/\phi_0)^{\sigma+1}]$, we find
that
\begin{equation}
{X\over M^4}={H^2M_{Pl}^4\over 128\pi^2\phi_0^2M^4
\left[1+(\sigma-1)L/8\pi G\phi_0^2\right]^{2\sigma\over\sigma-1}}~,
\end{equation}
both of which are proportional to $H^2$, but differ from $H^2/M^2$
in general.
Irrespective of the precise form of $V(\phi)$, Eq. (\ref{scalarflucts})
implies that in conventional (i.e. small $X/M^4$) slow roll cosmology
\begin{equation}
{X\over M^4}={H^4\over 8\pi^2M^4P_k^S}~.
\end{equation}
Thus, in models where $V=V_0[1+f(\phi/\phi_0)]$,
the slow roll approximation {\it and} the assumption of small
$X/M^4$ are only valid {\it simultaneously} provided 
\begin{equation}
{V_0\over M^2}\lesssim{3M_{Pl}^2(P_k^S)^{1/2}\over 2\sqrt{2}}
\approx 1.1\times 10^{-5}M_{Pl}^2(10^{10}P_k^S)^{1/2}.
\end{equation}
For $V_0=\epsilon M^4$, with $\epsilon\lesssim 1$, this condition implies 
$M\lesssim 0.003\epsilon^{-1/2}M_{Pl}$, or
a characteristic mass scale 
$M\lesssim 4\times 10^{16}\epsilon^{-1/2}$ GeV. In other
words, although small $X/M^4$ is more or less guaranteed by the slow
roll assumption provided $V_0\lesssim M^4$, the observational
constraint imposed by the magnitude
of $P_k^S$ restricts the values of $M$ for which the approximations
in conventional slow roll cosmology are all consistent.

The situation is a bit
trickier for models in which $V(\phi)$ is a powerlaw in $\phi$,
$V(\phi)=m^{4-n}\phi^n/n$, but with $n\neq 2$. (We also must
consider $n=4$ separately.) In this case, conventional inflation
yields
\begin{equation}
\dot\phi=-\sqrt{nm^{4-n}\over 24\pi G}\phi^{n/2-1}=
-M_{Pl}^{n/2}m^{2-n/2}\left({\phi\over M_{Pl}}\right)^{n/2-1}
\sqrt{n\over 24\pi},
\label{velnorm}
\end{equation}
and scalar fluctuations of amplitude
\begin{equation}
P^S_k={128\pi\over 3n^3}\left({\phi\over M_{Pl}}\right)^{n+2}
\left({M_{Pl}\over m}\right)^{n-4};
\label{flucnorm}
\end{equation}
the slow roll approximation requires
\begin{equation}
\left\vert{\dot\phi\over H\phi}\right\vert={n M_{Pl}^2\over
8\pi \phi^2}\ll 1,
\end{equation}
so that $\phi/M_{Pl}$ must be $\gtrsim 1$ in general. In fact,
the slow roll solution for $\phi$ is
\begin{equation}
{\phi\over M_{Pl}}=\sqrt{nL\over 4\pi}.
\label{phisol}
\end{equation}
From
Eq. (\ref{flucnorm}) we see that
\begin{equation}
{\phi\over M_{Pl}}=\left({3n^3P^S_k\over 128\pi}\right)^{1\over n+2}
\left({m\over M_{Pl}}\right)^{{n-4\over n+2}};
\end{equation}
since $P^S_k\sim 10^{-10}$, the first factor is small, so the
second must be large. Thus, for $n>4$, slow roll requires large
$m/M_{Pl}$ and for $n<4$, it requires small $m/M_{Pl}$. If we
regard the value of $P_k^S$ to be fixed observationally,
then we can solve Eq. (\ref{flucnorm}) to find
\begin{equation}
{M_{Pl}\over m}=\left({3n^3P^S_k\over 128\pi}\right)^{{1\over n-4}}
\left({\phi\over M_{Pl}}\right)^{-{n+2\over n-4}}
=\left({3n^3P^S_k\over 128\pi}\right)^{{1\over n-4}}
\left({4\pi\over nL}\right)^{{n+2\over 2(n-4)}},
\end{equation}
using Eq. (\ref{phisol});
combining with Eq. (\ref{velnorm}) implies
\begin{equation}
{X\over M^4}={n\over 48\pi}\left({m\over M}\right)^4
\left({3n^3P^S_k\over 128\pi}\right)^{{n\over n-4}}
\left({nL\over 4\pi}\right)^{-{4(n-1)\over n-4}}.
\label{condcheck}
\end{equation}
Eq. (\ref{condcheck}) shows that unless $m/M\ll 1$, $X/M^4\gg 1$
in conventional cosmology for all $n<4$, and that unless $m/M\gg 1$,
$X/M^4\ll 1$ for all $n>4$. For $n=4$, we take the potential to be
$V(\phi)=\lambda\phi^4/4$. In this case, Eq. (\ref{phisol}) continues
to hold true, and we find
\begin{eqnarray}
\dot\phi&=&-\sqrt{\lambda\over 6\pi}M_{Pl}\phi=
-\sqrt{\lambda L\over 6\pi^2}M_{Pl}^2\nonumber\\
P^S_k&=&{2\lambda L^3\over 3\pi^2}~.
\label{nfour}
\end{eqnarray}
Eqs. (\ref{nfour}) may be combined to find
\begin{equation}
{X\over M^4}={P_k^S\over 8L^2}\left({M_{Pl}\over M}\right)^4
\approx 3.5\times 10^{-15}(10^{10}P_k^S)\left({M_{Pl}\over M}\right)^4
\end{equation}
in this case, $X/M^4>1$ if $M_{Pl}/M>10^4$ or so, and vice-versa.

Thus we see, overall, that for potentials that are very flat, we do
not expect significant signatures of large mass scales in models
with $F(X)\neq X$, but for simple polynomial potentials $\propto
\phi^n$, such effects may be present, but are only likely for $n<4$. The
principal effect would be to change the dependence of the scalar
and tensor fluctuation amplitudes on the parameters of the theory.
This in turn would alter estimates of the energy scales associated
with inflation based on $P^S_k$ and $P^T_k$,
just as we found in \S~\ref{chaotic} for chaotic inflation.
Moreover, the ratio
$P_k^T/P_k^S$ is affected, but in a way that depends critically
on the form of $F(X)$, via $c_s$ (see Eqs.~(\ref{soundef}) and
(\ref{consistency})). Therefore, potentially, some information
about the scale of short distance physics
$M$ may be encoded
in the CMB fluctuations. Even though the determination of $M$ from
observations would be rather model-dependent, it may be possible,
at least in principle, to probe
short distance physics from cosmological measurements.

\acknowledgments

We thank the ITP, Santa
Barbara for hospitality while this work was initiated.
The work of G.S. was supported in part by the DOE grants
DE-FG02-95ER40893, DE-EY-76-02-3071, and the University of Pennsylvania
School of Arts and Sciences
Dean's fund. I.W. acknowledges support from NASA.
We are grateful to Henry Tye for useful comments on an
early draft of this paper.

\end{document}